\documentclass[11pt,a4paper]{article}
\usepackage[dvips]{graphicx}

\usepackage{epsfig}
\usepackage{epstopdf}
\usepackage{graphicx}
\usepackage{indentfirst}
\usepackage{amsmath}
\usepackage{amsfonts}
\usepackage{amssymb}
 \usepackage{textcomp}

\usepackage[pdftex,pdfmenubar=true,bookmarks=true,pdftoolbar=true]{hyperref}
\hypersetup{colorlinks,citecolor=blue,filecolor=magenta,linkcolor=magenta,urlcolor=cyan,pdftex}

\newcommand{\be}{\begin{equation}}
\newcommand{\ee}{\end{equation}}
\newcommand{\bea}{\begin{eqnarray}}
\newcommand{\eea}{\end{eqnarray}}

\begin{document}

\begin{center}
\begin{flushright}\begin{small}    
\end{small} \end{flushright} \vspace{1.5cm}
\Large{\bf Strong Magnetic field effects on Neutron Stars  within  $f(T)$ theory of gravity} 
\end{center}

\begin{center}
M. G. Ganiou $^{(a)}$\footnote{e-mail:moussiliou\_ganiou @yahoo.fr},
C. A\"{i}namon   $^{(a)}$\footnote{e-mail: ainamoncyrille@yahoo.fr},
M. J. S. Houndjo   $^{(a,b)}$\footnote{ e-mail: sthoundjo@yahoo.fr}
and J. Tossa $^{(a)}$\footnote{e-mail: joel.tassa@imsp-uac.org}
\vskip 4mm
$^a$ \,{\it Institut de Math\'{e}matiques et de Sciences Physiques (IMSP)}\\
 {\it 01 BP 613,  Porto-Novo, B\'{e}nin}\\
  $^{b}$\,{\it Facult\'e des Sciences et Techniques de Natitingou - Universit\'e de Parakou - B\'enin} \\
\vskip 2mm
\end{center}

\begin{abstract}
\hspace{0,2cm} 
We investigate in this paper the structures of neutron stars under the strong magnetic field in the framework of $f(T)$ gravity where $T$ 
denotes the scalar torsion. The TOV equations in this theory of gravity have been considered
and numerical resolution of these equations has been performed 
within  perturbative approach 
taking into account  the equation of state of neutron dense matter in magnetic field. 
We simplify the problem by considering
the very strong magnetic field which  affects considerably the dense matter;
and for quadratic and cubic corrections  to Teleparallel  term,
one finds  that the mass of neutron stars can increase for different values of the 
perturbation parameter. The deviation from Teleparallel for different values of 
 magnetic field is found 
out and this feature is very appreciable in the case of cubic correction.
Our results are related to the hadronic particles description
with very 
small hyperon contributions and the mass-radius evolution is consistency with  
the observational data. 
\end{abstract}


\tableofcontents

\section{Introduction}
The explanation of the present state of our Universe and the several structures appearing 
in it, has opened the gate to  various reflexions.
Indeed,  the current acceleration of the Universe is widely accepted through several  
independent observational
data, as supernovae Ia \cite{Eddington}-\cite{Riess}, the large scale structure of the Universe \cite{Brax}, 
cosmic shear through gravitational weak lensing surveys \cite{Schimdt}  
and the Lyman alpha forest absorption lines \cite{Spergel}. 
Due to the need to understand the acceleration of the Universe, various 
theories have been introduced. 
The well known standard equivalent theories, the General Relativity (GR)
and the Teleparallel Theory, are the first theories used for explaining the acceleration of the Universe,
including the existence of the dark energy as a new component of the Universe \cite{capo}. 
We recall here that the Teleparallel Theory is 
an idea of Einstein in 1928 in his attempt to unify  gravity and electromagnetism. 
This theory nowadays, is formulated as 
higher gauge theory \cite{ Baez}; furthermore it has already been established 
that GR can in fact be re-cast into Teleparallel
language \cite{So}, known as the Teleparallel Equivalent of General Relativity (TEGR).
The  second attempt in  theories of gravity  for explaining 
this state of Universe consists to modify the  GR and the Teleparallel Theory. 
Due to the fact that the  Teleparallel theory does not yield consistent results according 
to the observational data, the immediate attempt to modify  this theory of gravity 
is replacing the Teleparallel term, the torsion scalar $T$ by an algebraic function 
$f(T)$ based on the Weitzenb\"{o}ck connection instead of Levi-Civita connection used
in GR. \par
 Furthermore, structure formations have also been investigated in this modified 
 theory of gravity. 
 The  neutron stars  structures
  have already been executed in several theories of gravity  mainly in
  $f(R)$ \cite{Artyom,Artyom2,Capozziello} and $f(T)$  \cite{Houndjo}. Indeed 
  the structures of neutron and quark Stars have recently been investigated by \cite{Houndjo} 
  through the deviation of the mass-radius diagrams for power-law and exponential 
  type correction from the Teleparallel Theory gravity by considering some equations of state.  
  In this paper, we focus our attention on the study of these stars in particular 
  the neutron stars by investigating the effects of  strong magnetic fields 
 in the framework of $f(T)$ gravity. Note that such studies have been performed in the so-called 
 $f(R)$ theory of gravity where interesting result  have been found 
 \cite{Artyom,Rabhi2,Arapoglu,Capozziello,Astashenok, Rabhi}).
We emphasize here that 
neutron stars are observed as several classes of self-gravitating systems \cite{Artyom}. The
structure of neutron stars and the relation between the mass and the radius are determined by equations of state
(EOS) of dense matter. Moreover, the maximal mass of neutron star is still an open question.
Recent observations allow 
to estimate this limit at least as $2M_\odot$. It exits some well-measured limits 
according to 
several observational data on the pulsars and  massive neutron stars:  there are
   $1.972M_\odot $  for pulsars PSR J$1614-2230$ \cite{Demorest}, 
   $2.01M_\odot $  for pulsars J$0348+0432$ \cite{ Antoniadis}, 
  $1.972M_\odot $ for  Vela X-$1$ \cite{Rawls}, $2M_\odot $ for  $4$U$1822-371$\cite{Darias},   
  $2.7M_\odot $ for  J$1748-2021$B \cite{Freire}. 
 Several works with different approaches show that  the mass limit of  neutron star 
 can increase: larger hyperon-vector couplings with 
 stiffness of the EOS \cite{Hofmann},  model with chiral quark-meson coupling \cite{Miyatsu},
 
 the quark-meson coupling model which naturally incorporates hyperons \cite{Whittenbury}
 and others.
However, it has been shown in the literature that ultrastrong magnetic field affected 
considerably  the equation of
state (EOS) of neutron-star matter  and  consequently  could lead to the increase of 
the maximal mass limit of these stars \cite{Broderick}. 
In order to explain this fact, various  models of dense nuclear matter 
in presence of strong magnetic fields have been considered  such as models 
with hyperons and quarks, and model with interacting $npe\mu$ gas. 
It has also been performed that the Landau quantization leads to the softening of the EoS for matter 
but account for contributions of magnetic field into pressure and density.\par
In this paper,  we present the models of neutron star for simple  EOS in the ultrastrong magnetic field via the $f(T)$ gravity. 
In Ref.\cite{Broderick}, 
it  is clearly shown that  the structure of a magnetized neutron
star will be mostly affected by contributions from the magnetic field stress 
$P_f = B^2 /8\pi = 4.814 \times 10^{-8}B^{*2}$ Mev f$\text{m}^{-3}$ 
which greatly exceeds the matter pressure $P_m$ at all relevant densities
for $B^{*}\geq 10^{5}$. Having in hand such fundamental result and through
a perturbative approach, one obtains that the maximal mass limit of neutron stars
can increase under an ultrastrong field magnetic. We 
consider two corrections of the Teleparallel Theory (quadratic and cubic corrections) 
and find that the parameter of correction plays an important role in this analysis. 
The deviation of mass-radius from Teleparallel Theory
 appears clearly in the case of cubic corrections.  
 Our investigation through perturbative approach shows  that in the case  of
 null magnetic  field, the 
 mass of neutron stars can increase as it was predicted by \cite{Artyom}. 
 The paper is organized as follows. In Sec.II, we make some reviews on 
 the Teleparallel Theory via its equivalence to GR and its modified version, 
 the so-called $f(T)$ gravity. In Sec.III, we 
 present the TOV equations and its perturbative versions.
 Strong magnetic field effect on the dense matter  in the framework of the relativistic 
 mean field has been described in the Sec.IV. Our main results have been presented 
 in  the Sec.V and the conclusions in Sec.VI.

\section{From Teleparallel equivalent of General Relativity to f(T)  }
 In  Teleparall, equivalent of General Relativity, as in  General Relativity, the structure of space-time  is represented by a manifold $ M$.
 At
every point $p\in M$ in the local coordinate chart $\{ x^1,...,x^n\}$, the tangent space $T_pM$ at $p$
is spanned by the coordinate vector fields $\{ \partial_1,...,\partial_n\}$. The corresponding dual space is denoted by  $T^*_pM$  and 
generated by $\{ dx^1,...,dx^n\}$. In addition, the tangent space is a $4$
dimension space described by  the Lorentzian metric $g$ with
signature $(-,+,+,+)$. We will label all space-time coordinates
by Greek subscripts that run from $0$ to $3$, with $0$ denoting the time
dimension, while all spatial
coordinates will be labeled by $i$, $j$, $k$, . . . that run 
from $1$ to $3$.\par 
Let us assume $\{e_A(x)\}$ as an arbitrary base of $T_pM$. 
We can express the total derivative covariant as $\nabla e_A$:
\begin{equation}
  \nabla e_A(x)=\Gamma^B_{\;A}(x)e_B(x),
 \end{equation}
 with $\Gamma^B_{\;A}(x)$, the  1-form connection satisfying
 \begin{equation}
  \Gamma^B_{\;A}(x)=\langle z^B(x), \nabla e_A(x) \rangle= \Gamma^B_{\;\nu A}(x)dx^\nu. 
 \end{equation}
 In this last expression, $\{z^B(x)\}$ represents the dual 
 of $\{e_B(x)\}$ where as $\{dx^\nu\}$ stayed for base dual of local base 
 $\{\partial_\nu\}$. 
 Then, it comes that  $e_A=e^\nu_{\;A}\partial_\nu$ and  $z^B=e^B_{\;\;\nu}dx^\nu$. 
 In this paper, the capital letters 
 $A$, $B$, $C$,...  take the values running from $0$ to $3$ and  $a$, $b$, $c$... 
 take the values  running from $1$ to $3$. \par
 Assuming that the spacetime is parallelisable
 (i.e. there exist $n$ vector fields $\{v_1,...,v_n \}$ such
that at any point  $p\in M$ the tangent vectors $v_i|_p$
provide a basis of the tangent space at
$p$), the mapping between the bases in coordinate frame $\{\partial_\mu\}$ 
to that of non-coordinate frame
$\{e_A\}$ is an isomorphism $TM\rightarrow M\times \mathbf{R}^4$. 
This also comes from  the fact that all $n$  
dimensional parallelisable manifold  has a tangent space which can
be decomposed in the direct product of  $M$ and $\mathbf{R}^n$. 
 Note that the frame field depends only on the affine structure of
 the manifold and hence {\it a priori} has any relation with the metric.
Actually, we describe the frames with  the metric  by equipping the $\mathbf{R}^4$ 
space with the minkowskian metric 
$\eta_{AB}$ such as  
 \begin{equation}
  g_{\mu\nu}= \eta_{AB}e^A_{\;\;\mu}e^B_{\;\;\nu}.
 \end{equation}
 This also means  that on the manifold $M$,  we arbitrary choose  a frame $\{e_A\}$
 at each point namely locally 
 on some open chart  $U\subset M$. This approach can be extended  
 by parallelisability. One can define the metric $\eta$ 
 on the open chart $U$ by 
 
 \begin{equation}
  \eta(e_A,e_B)= \eta_{AB},
 \end{equation}
 which shows the orthonormality on the tetrad. 
 It is important to notify here that for the $(3+1)$ dimension gravity, the previous results
  hold according to the  Steenrod theorem: all $3$ dimensional
  and orientable  manifold is parallelisable \cite{Steenrod, Stiefel}.
 Thus
a $4$-dimensional spacetime with orientable spatial section is also parallelisable. 
To put it
slightly differently, if any spatial slice of spacetime is an orientable $3$-manifold 
(and as such
parallelizable) with initial data that can be propagated uniquely in time, in the manner of
$3+1$ decomposition of $ADM$ \cite{Arnowitt}, 
then the entire spacetime is parallelizable. We also emphasize here from 
Geroch theorem \cite{Geroch} that a non-compact 4-dimensional Lorentzian manifold M admits a spin structure if and only
if it is parallelisable. \\
Let introduce  $\stackrel{\bullet}{\nabla}$, the  Weitzenb\"{o}ck connection  \cite{ Weitzenbock} defined by
 \begin{equation}
 \stackrel{\bullet}{\nabla}_XY:=(XY^A)e_A, 
 \end{equation}
with $Y=Y^Ae_A$. The teleparallelism condition on the tetrads 
imposes $\stackrel{\bullet}{\nabla}e_A=0$ 
which allows to define the coefficients of the connection as 
\begin{equation}\label{con}
 \stackrel{\bullet}{\Gamma}_{\;\;\;\mu\nu}^{\;\,\lambda}= e_A^{\;\;\lambda}\partial_\nu e^A_{\;\;\mu}
 =-e^A_{\;\;\mu}\partial_\nu e_A^{\;\;\lambda}.
 \end{equation} 
 The coefficients  defined in (\ref{con}) is for  an unique connection 
 \cite{Youssef} which, at each vector field $ X $, 
 gives rise to parallelisation on $M$ , assuming of course  that $M$ is parallelisable.

\begin{equation}
 \stackrel{\bullet}{T}(X,Y)=\stackrel{\bullet}{\nabla}_XY- \stackrel{\bullet}{\nabla}_YX-[X,Y].  
\end{equation}
And then  
\begin{equation}\label{tor}
  \stackrel{\bullet}{T}(X,Y)= X^AY^B[e_A,e_B].
\end{equation}

 The equation (\ref{tor}) shows that the torsion tensor associated  to   
 Weitzenb\"{o}ck connection 
is   generically non zero  because in general,  the 
   basis vectors $ \{e_A \} $ are  not integrable.
   In the local coordinates, the torsion tensor components can be expressed by
  \begin{equation}
   \stackrel{\bullet}{T}_{\;\;\;\mu\nu}^{\;\,\lambda}= \stackrel{\bullet}{\Gamma}_{\;\;\;\nu\mu}^{\;\,\lambda}-
   \stackrel{\bullet}{\Gamma}_{\;\;\;\mu\nu}^{\;\,\lambda}=e_A^{\;\;\lambda}(\partial_\mu e^A_{\;\;\nu}-\partial_\nu e^A_{\;\;\mu})\neq 0.
  \end{equation}
  We also define here the curvature tensor associated to 
  the Weitzenb\"{o}ck connection through  the Riemann
curvature by 
 \begin{equation}\label{riem}
  \stackrel{\bullet}{R}(X,Y)Z=\Big( \stackrel{\bullet}{\nabla}_X \stackrel{\bullet}{\nabla}_Y-\stackrel{\bullet}{\nabla}_Y
  \stackrel{\bullet}{\nabla}_X-\stackrel{\bullet}{\nabla}_{[X,Y]}\Big)Z.
 \end{equation}
From the fact that  $\stackrel{\bullet}{\nabla}_Xe_A=(X\delta^C_A)e_c=0$, one gets  
 \begin{equation}
  \stackrel{\bullet}{R}(e_A,e_B)e_C=\stackrel{\bullet}{\nabla}_{e_A}(\stackrel{\bullet}{\nabla}_{e_B}e_C)-
  \stackrel{\bullet}{\nabla}_{e_B}(\stackrel{\bullet}{\nabla}_{e_A}e_C)-\stackrel{\bullet}{\nabla}_{[e_A,e_B]}e_C=0.
 \end{equation}
It follows that the curvature tensor associated to the Weitzenb\"{o}ck connection 
is equal to zero contrary to the case of the
Levi-Civita connection in GR.  
We can thus say that the curvature is an intrinsic property  of spacetime but  it
also  only depends on the connection that is put on spacetime.\par
Another important tensor emerging from the use of the Weitzenb\"{o}ck connection 
is the contortion which shows the difference between 
the  Weitzenb\"{o}ck connection and the Levi-Civita  connection \cite{equiv} according to
\begin{center}
\begin{equation}\label{cont}
  \stackrel{\bullet}{K}_{\;\;\;\mu\nu}^{\;\,\lambda}:= \stackrel{\bullet}{\Gamma}_{\;\;\;\mu\nu}^{\;\,\lambda}-
  \Gamma_{\;\;\;\mu\nu}^{\;\,\lambda}=\frac{1}{2}\Big(\stackrel{\bullet}{T}_{\nu}{}^{\lambda}{}_{\mu}+
  \stackrel{\bullet}{T}_{\mu}{}^{\lambda}{}_{\nu}- \stackrel{\bullet}{T}^{\lambda}{}_{\mu\nu}\Big),
\end{equation}
\end{center}
or in the equivalent form as  
\begin{center}
 \begin{equation}
 \stackrel{\bullet}{K}{}^{\mu\nu}{}_{\beta}=-\frac{1}{2}\Big( \stackrel{\bullet}{T}{}^{\mu\nu}{}_{\beta} -
\stackrel{\bullet}{T}{}^{\nu\mu}{}_{\beta}-\stackrel{\bullet}{T}_{\beta}{}^{\mu\nu}{} \Big),      
 \end{equation}
 \end{center}
 where  $ \Gamma_{\;\;\;\mu\nu}^{\;\,\lambda}$  represents the coefficient of  
 Levi-Civita connection.
 In order to show that  the both
  connections make  the same description of spacetime namely, 
  prove that the  Teleparallel Theory based on the Weitzenb\"{o}ck connection
   is equivalent to GR using the Levi-Civita connection,
   we recall here the action of Einstein-Hilbert
   GR as
 \begin{equation}\label{act}
  S=\frac{1}{2\kappa^2}\int d^4x\sqrt{-g}R, 
 \end{equation}
 With $\kappa^2=\frac{16\pi G}{c^4}$, and  $R$
 is the Ricci scalar curvature coming from the Levi-Civita  connection. 
 The action  (\ref{act})
 can be written with scalar torsion as  
 \begin{equation}\label{act1}
  S=\frac{1}{2\kappa^2}\int d^4x eT, 
 \end{equation}
 where $e=|\text{det}(e^A{}_\mu)|$ is equivalent to 
 $\sqrt{-g}$ in  GR and  $T$ the scalar  torsion defined by  
 \begin{equation}\label{de}
  \stackrel{\bullet}{T}:=\stackrel{\bullet}{S}_\beta{}^{\mu\nu}
  \stackrel{\bullet}{T}{}^\beta{}_{\mu\nu},
 \end{equation}
with 
\begin{equation}\label{sup}
 \stackrel{\bullet}{S}_\beta{}^{\mu\nu}=\frac{1}{2}\Big(  \stackrel{\bullet}{K}{}^{\mu\nu}{}_{\beta}+
 \delta^\mu_\beta \stackrel{\bullet}{T}{}^{\alpha\nu}{}_{\alpha}- \delta^\nu_\beta \stackrel{\bullet}{T}{}^{\alpha\mu}{}_{\alpha}  \Big).
\end{equation}
By using the relations (\ref{cont}) and  (\ref{sup}), the scalar torsion defined in (\ref{de}) can be explicitly put in 
the following relations 
\begin{equation} 
 \stackrel{\bullet}{T}= \frac{1}{4}\stackrel{\bullet}{T}{}^\beta{}_{\mu\nu}\stackrel{\bullet}{T}{}_\beta{}^{\mu\nu} +
 \frac{1}{2}\stackrel{\bullet}{T}{}^\beta{}_{\mu\nu}\stackrel{\bullet}{T}{}^{ \nu\mu}{}_\beta-
\stackrel{\bullet}{T}_{\beta\nu}{}^\beta\stackrel{\bullet}{T}{}^{ \mu\nu}{}_\mu. 
\end{equation} 
 In the local base $\{\partial_\mu \}$, the components of Riemann tensor given in (\ref{riem}) can be obtained in the framework of 
 Levi-Civita connection  by
\begin{equation}
 R^\rho{}_{\mu\lambda\nu}=\partial_\lambda\Gamma^\rho{}_{\mu\nu}-\partial_\nu\Gamma^\rho{}_{\mu\lambda}+
 \Gamma^\rho{}_{\sigma\lambda}\Gamma^\sigma{}_{\mu\nu}-\Gamma^\rho{}_{\sigma\nu}\Gamma^\sigma{}_{\mu\lambda}.
\end{equation}
By making using the relation (\ref{cont}) and after some contractions, one obtains the  associated Ricci tensor as  
\begin{equation}
 R_{\mu\nu}= \nabla_\nu\stackrel{\bullet}{K}{}^{\rho}{}_{\mu\rho}-\nabla_\rho\stackrel{\bullet}{K}{}^{\rho}{}_{\mu\nu}+
 \stackrel{\bullet}{K}{}^{\rho}{}_{\sigma\nu}\stackrel{\bullet}{K}{}^{\sigma}{}_{\mu\rho}-
 \stackrel{\bullet}{K}{}^{\rho}{}_{\sigma\rho}\stackrel{\bullet}{K}{}^{\sigma}{}_{\mu\nu},
\end{equation} 
where $\nabla$ represents the  covariant derivative in GR. By  combining the relation (\ref{cont}) with the following relations
$\stackrel{\bullet}{K}{}^{(\mu\nu)\sigma}=\stackrel{\bullet}{T}{}^{\mu(\nu\sigma)}=
\stackrel{\bullet}{S}{}^{\mu(\nu\sigma)}=0$ 
and by taking into consideration 
\newline
$ \stackrel{\bullet}{S}{}^\mu{}_{\rho\mu}=
2\stackrel{\bullet}{K}{}^\mu{}_{\rho\mu}=-2\stackrel{\bullet}{T}{}^\mu{}_{\rho\mu}$, we have 
\begin{equation}\label{ric}
 R_{\mu\nu}=-\nabla^\rho \stackrel{\bullet}{S}_{\nu\rho\mu}-g_{\mu\nu}\nabla^\rho\stackrel{\bullet}{T}{}^{\sigma}{}_{\rho\sigma}- 
 \stackrel{\bullet}{S}{}^{\rho\sigma}{}_{\mu}\stackrel{\bullet}{K}_{\sigma\rho\nu}, 
\end{equation}
whose total contraction gives 
  \begin{equation}\label{sca}
     R=-\stackrel{\bullet}{T} -2\nabla^{\mu}\stackrel{\bullet}{T}{}^\nu{}_{ \mu\nu}.
  \end{equation}
This last relation finds out the trivial equivalence of GR and Teleparallel under the actions defined in 
(\ref{act})  and (\ref{act1}). It follows that these two relations are the  Einstein-Hilbert action respectively in G and 
Teleparallel. In the rest  of this work, we will leave the   strackrel bullet ($\bullet$)   on all quantities  resulting from 
the Weitzenb\"{o}ck connection.\par
The action of the modified versions of  TEGR (Teleparallel Equivalent of GR) is obtained by substituting the scalar torsion 
of the action (\ref{act1}) by an arbitrary function of scalar torsion giving then to modified theory $f(T)$. 
This approach is similar in spirit to the generalisation of 
Ricci scalar curvature of Einstein-Hilbert action (\ref{act}) ) by a function of this scalar  leading  to the well known 
$ f (R) $ theory. The action of $f(T)$ theory can be defined  as 
\begin{eqnarray}\label{action}
 S=\frac{1}{2\kappa^2}\int  \left[f(T)+\mathcal{L}_{Matter}\right]ed^4x,
\end{eqnarray} 
with $T$  the scalar torsion and  $\mathcal{L}_{Matter}$  
the matter density Lagrangian which only depends from the  
tetrads $e_A{}^{\mu}$ without their derivative. 
The variation of the action with respect to tetrads $e_A{}^{\mu}$ gives 
\cite{salako, Houndjo}
 \begin{eqnarray}\label{mot}
  \frac{1}{e}\partial_\mu(eS_A{}^{\mu\nu})f_T(T)- e_A{}^{\lambda}T^\rho{}_{\mu\lambda}S_\rho{}^{\mu\nu}f_T(T) 
  +S_A{}^{\mu\nu}\partial_\mu(T)f_{TT}(T)+\frac{1}{4} e_A{}^{\nu}f(T)= \Theta^\nu_A, 
 \end{eqnarray}
 with $f_T(T)=df(T)/dT$,   $f_{TT}(T)=d^2f(T)/dT^2$ and   
 $\Theta^\nu_A$ the energy-momentum tensor naturally related to the
 matter.\par 
 \section{The Generalised Tolman-Oppenheimer-Volkoff (TOV) 
 equations in $f(T)$ } 
 TOV equations are often used in gravitational theories 
 to study the structure of relativistic stars. 
 Their determination requires the choice  of a given 
 theory of gravity, the precision on the type of spacetime (metric) 
 and the matter  source in the Universe. In the 
 framework of our present work, based on  the $ f (T) $ theory,
we express the geometric Lagrangian density as: $ f (T) = T + \xi g (T) $
with $ \xi $ a real constant such that $ \xi=0 $  matches
  to the Teleparallel  theory. The equations of motion (\ref{mot}) become
\cite{Houndjo}

\begin{eqnarray}\label{mot1}
  [\frac{1}{e}e^A{}_\sigma \partial_\mu(ee_A{}^{\rho}S_\rho{}^{\nu\mu})+ T^\rho{}_{\mu\sigma}S_\rho{}^{\mu\nu}](1+\xi g_T(T)) 
  +\xi S_\sigma{}^{\nu\mu}\partial_\mu(T)g_{TT}(T)+\frac{1}{4}\delta_\sigma^{\nu}(T+\xi g(T))= \frac{\kappa^2}{2}\Theta^\nu_\sigma, 
 \end{eqnarray}
 In order to obtain solutions that describe the stellar objects, 
 we consider a spherical symmetric metric
  having two functions $\varphi$ and  $\lambda$ depending 
  on the radial coordinate by
 \begin{eqnarray}\label{metrique}
  ds^2= -e^{2\varphi}dt^2+ e^{2\lambda}dr^2+r^2(d\theta^{2}+\sin^{2}\theta
  d\phi^{2}),
\end{eqnarray}
which can be generated by the following table of tetrads
\begin{eqnarray}
 e^A{}_{\mu}=\text{diag}(e^{\varphi}, e^{\lambda},r, r\sin\theta ).
 \end{eqnarray}  
Let recall here that the most successful  metric in the stellar objets
description is the  one of Schwarschild \cite{Houndjo,Artyom,Stephani,Cooney} 
which is  a particular case of metric
described in (\ref{metrique}) with the following fundamental  relation
\begin{eqnarray}\label{fondamentale}
 e^{-2\lambda}=1-\frac{2GM}{c^2r}.
\end{eqnarray}

 We also consider that the  Universe described  by this metric has
 as content, an isotropic fluid such that the 
 associated energy-momentum tensor
  can be expressed by
 \begin{eqnarray}\label{ener imp}
  \Theta^\nu_\sigma=-(\rho+P)u_\sigma u^\nu+P\delta^\nu_\sigma,
 \end{eqnarray}
with   $u^\nu$ the four-velocity, 
$\rho$ and $P$ the energy density and the pressure 
of the matter respectively.  Having the metric
(\ref{metrique}) and the energy-momentum (\ref{ener imp}) 
in hand, we  establish the field equations from the
general equation  (\ref{mot1})
via the following relations 
\begin{eqnarray}
 \label{friedman equation 1}
  \frac{1}{4}[T+\xi g(T)]-\frac{1}{2}[T-\frac{1}{r^2}-2\frac{e^{-2\lambda}}{r}(\varphi^\prime+\lambda^\prime)](1+\xi g_T(T))
  &=&-\frac{4\pi\rho}{c^4},\\
  \label{friedman equation 2}
  -\frac{1}{4}[T+\xi g(T)]-\frac{1}{4}\left[\frac{3T}{2}-\frac{1}{r^2}+e^{-2\lambda}[\varphi^{\prime\prime}+(\varphi^\prime+
  \frac{1}{r})(\varphi^\prime-\lambda^\prime)]\right](1+\xi g_T(T))&=&\frac{4\pi P}{c^4},\\
  \frac{\cot \theta}{2r^2}\xi T^\prime g_{TT}(T)&=&0.
 \end{eqnarray}
 The ``prime" in these relations means the derivative with respect to the 
 radial coordinate $r$.
 The scalar torsion can also be calculated as
 \begin{eqnarray}\label{torsion}
  T=-\frac{2e^{-2\lambda}}{r^2}(2r\varphi^\prime+1). 
 \end{eqnarray}
However, from the conservation of the energy-momentum tensor,   
$\bigtriangledown_\mu\Theta^\mu_\sigma=0$, the quantity  $\varphi^\prime$ 
can be expressed in term of the  energy density and the pressure as 
 \begin{eqnarray}\label{varphi}
  \varphi^\prime=-(\rho+P)^{-1}\frac{dP}{dr}.
 \end{eqnarray}
In order to obtain the TOV equations,
we introduce the following dimensionless variables   $M=mM_\odot$, 
$r\rightarrow r_g r$, $\rho \rightarrow \rho M_\odot/r_g^3$, 
$P\rightarrow p M_\odot c^2/r_g^3$, $T\rightarrow T/r_g^2$, 
  $\xi r_g^2 g(T)\rightarrow \xi g(T)$, 
  with $r_g=GM_\odot/c^2=1.47473\text{km}$. 
  Indeed, the relation  (\ref{fondamentale}  becomes
  
\begin{eqnarray}\label{fondamentale}
 e^{-2\lambda}=1-\frac{2m}{r},
\end{eqnarray}
 and 
 \begin{eqnarray}\label{lambda}
  \lambda^\prime=\left(\frac{m^\prime}{r}-\frac{m}{r^2}\right)\left(1-\frac{2m}{r}\right)^{-1}.
 \end{eqnarray} 
 The scalar torsion becomes  (\ref{torsion}) 
  \begin{eqnarray}\label{new torsion}
   T(r)= -\frac{2}{r_g^2r^2}(1-\frac{2m}{r})\left[1-2rr_g(\rho+p)^{-1}\frac{dp}{dr}\right].
  \end{eqnarray}
By using the field  equations (\ref{friedman equation 1}),
(\ref{friedman equation 2}) and the relations 
(\ref{varphi}), (\ref{lambda}) and  (\ref{new torsion}), 
one gets the TOV equations as
 \begin{eqnarray}
 \label{TOV1}
 -4\pi\rho &=&-(1-\frac{2m}{r})\Bigg[\frac{1}{2r_g^2r^2}\left(1+2\xi g_T(T)\right)+\frac{1}{2r_gr}\left( 1-r_g^2+2\xi g_T(T)-
  r_g^2\xi g_T(T)\right)(\rho+p)^{-1}\frac{dp}{dr}+\\ \nonumber
  &&+\frac{r_g}{2r^2}\left(1+2\xi g_T(T)\right)\Bigg]
  -\frac{1}{2r^2}\left(1+\xi g_T(T)-r_g-r_g\xi g_T(T)\right)+\frac{1}{4}\xi g(T)
  +\frac{r_g}{r^2}\left(1+\xi g_T(T)\right)\frac{dm}{dr},\\
  \label{TOV2}
 4\pi p &=&-(1-\frac{2m}{r})\Bigg\{ -\frac{1}{2r_g^2r^2}+\frac{1}{2r_gr} (\rho+p)^{-1}\frac{dp}{dr}+\Big[ -\frac{3}{2r_g^2r^2}\\\nonumber
 &&+\frac{1}{2r_gr} (\rho+p)^{-1}\frac{dp}{dr}+r_g^2\frac{d}{dr}[ (\rho+p)^{-1}\frac{dp}{dr}]-r_g^2(\rho+p)^{-2}(\frac{dp}{dr})^2\\\nonumber
 &&+ \frac{r_g}{2r}(\rho+p)^{-1}\frac{dp}{dr}-\frac{r_g^2}{2r}(\rho+p)^{-1}\frac{dp}{dr}+\frac{r_g}{4r^2}\Big] 
 \left(\frac{1+\xi g_T(T)}{4}\right)  \Bigg\}- (\frac{1+\xi g_T(T)}{4})\\\nonumber
 &&\times[-\frac{r_g^2}{2r}(\rho+p)^{-1}\frac{dp}{dr}
  +\frac{r_g}{4r^2}]-\frac{1}{4}\xi g(T)+ \frac{1}{4r^2}\left(1+\xi g_T(T)\right)+\\\nonumber
  &&+\frac{r_g^2}{r}\Big[-(\rho+p)^{-1}\frac{dp}{dr}+
  \frac{1}{2r_gr}\Big]\left(\frac{1+\xi g_T(T)}{4}\right) \frac{dm}{dr},\\
  \label{trace equation}
  4\pi(\rho-3p)&=& -(1-\frac{2m}{r})\Bigg\{  \frac{1}{4r_g^2r^2} (17+13\xi g_T(T))+  
  \frac{1}{r_gr} \left( -\frac{13}{4}-r_g^2-\frac{5}{2}\xi g_T(T)-
  r_g^2\xi g_T(T)\right) \\\nonumber
  &&\times(\rho+p)^{-1}\frac{dp}{dr}+\frac{r_g}{2r^2}\left(1+2\xi g_T(T)\right)  +  
  \Big[  3r_g^2(\rho+p)^{-2}(\frac{dp}{dr})^2-3r_g^2\frac{d}{dr}[ (\rho+p)^{-1}\frac{dp}{dr}] \\\nonumber
  &&+\frac{3r_g}{2r}(r_g-1)(\rho+p)^{-1}\frac{dp}{dr}-\frac{3r_g}{4r^2}\Big]\left(\frac{1+\xi g_T(T)}{4}\right)\Bigg\}
   -\left( \frac{11r_g+4}{4r^2} +  \frac{3r_g^2}{2r} (\rho+p)^{-1}\frac{dp}{dr}\right) \\\nonumber 
   && \times\left(\frac{1+\xi g_T(T)}{4}\right)
  +\xi g(T)+\Big[\frac{5r_g}{2r^2}+\frac{3r_g^2}{r}(\rho+p)^{-1}\frac{dp}{dr}\Big]\left(\frac{1+\xi g_T(T)}{4}\right)\frac{dm}{dr}.
 \end{eqnarray} 
  The last equation, namely the equation 
  (\ref{trace equation}), is the trace of
  the equations (\ref{friedman equation 1}) and  
  (\ref{friedman equation 2}). We emphasize 
  here that in order to obtain the stellar structures, 
  it is important to require  some asymptotic
  flatness as the radial coordinate evolves \cite{Houndjo}.
  
  \begin{eqnarray}
  \mbox{lim}_{r \rightarrow +\infty} T(r) =0, \qquad   \mbox{lim}_{r \rightarrow +\infty} m(r) =\text{cst}
 \end{eqnarray}
The relations (\ref{TOV1}),  (\ref{TOV2}) and   (\ref{trace equation}) 
can be numerically solved by having  the  algebraic function 
$g(T)$ . But in this work, one can use the 
perturbative approach. In this approach, the terms containing $g(T)$ 
must be of first order of the small parameter $\xi$. 
\subsubsection{Perturbation TOV equations}
In order to  solve numerically the equations 
 (\ref{TOV1}), (\ref{TOV2}) and (\ref{trace equation}), 
one proceeds by  a perturbative approach 
. In the framework of a perturbative
solution, the density, pressure, mass and  scalar 
torsion can be expanded as \cite{Arapoglu,Capozziello, Astashenok}   
 \begin{eqnarray}
  p=p^{(0)}+\xi p^{(0)},  \qquad \rho=\rho^{(0)}+\xi \rho^{(0)},\\
  m=m^{(0)}+\xi m^{(0)},   \qquad T=T^{(0)}+\xi T^{(0)},
 \end{eqnarray}
where $p^{(0)}$, $\rho^{(0)}$, $m^{(0)}$ and $T^{(0)}$ 
satisfy the standard TOV equations 
(see \cite{Arapoglu} for the zeroth order mass).  
The scalar torsion at zeroth order can be specified 
as $T^{(0)}=-16\pi \rho^{(0)}$.
Finally, perturbative TOV equations can be established by
 \cite{Astashenok}  
 \begin{eqnarray}
  \frac{r_g}{r^2}\frac{dm}{dr}&=&\xi g^{(0)}_T(T)\Big[\frac{1}{2r^2}(1-r_g)+\frac{1}{r^2r^2_g} 
  (1-\frac{2m^{(0)}}{r})-8\pi r_g\rho^{(0)}\Big]+\frac{1}{2r^2}(1-r_g)+\\\nonumber
 &&+\frac{1}{2r^2r^2_g} (1-\frac{2m}{r})-\frac{1}{r_g r}(1-r^2_g)(\rho+p)^{-1}\frac{dp}{dr}-\frac{1}{4}\xi g^{(0)}(T)-4\pi \rho,
 \end{eqnarray}
\begin{eqnarray}
 (\rho+p)^{-1}\frac{dp}{dr}\Big[ -\frac{r^2_g}{4r} \frac{dm}{dr}+\frac{r^2_g}{8r} - 
 (1-\frac{2m}{r})\Big(  \frac{5}{8r_g r} - \frac{3}{r^2r^2_g}+\frac{1}{4}r_g^2+\frac{r_g}{8r}-    \frac{r_g^2}{8r}  \Big)  \Big]-\\\nonumber
 -\frac{1}{4}(1-\frac{2m}{r})\Big[  r_g^2\frac{d}{dr}[ (\rho+p)^{-1}\frac{dp}{dr}]  -r_g^2(\rho+p)^{-2}(\frac{dp}{dr})^2+\frac{r_g}{4r^2}
 -\frac{2}{r^2r^2_g}\Big]+\\\nonumber
 +\xi \Big[ -\frac{1}{4}(1-\frac{2m^{(0)}}{r}) 
 (-\frac{3}{2r^2r^2_g}+\frac{r_g}{4r^2})
 +\frac{1}{4r^2}+\pi\rho^{(0)}r_g\Big]g^{(0)}_T(T)
 -\frac{1}{4}\xi g^{(0)}(T)-\frac{r_g}{16r^2}-\frac{1}{4r^2}-4\pi p=0.
 \end{eqnarray}

 \section{Strong magnetic field effect on the dense matter  
 in the framework of the relativistic mean field: brief reviews }  
 Our goal in this work consists to study the effect of strong magnetic
 field on the neutron stars in the framework of $f(T)$
  theory. In general, for nuclear matter containing baryon octet  
  $(b=p, n, \Lambda,\Sigma^{0,\pm},\Xi^{0,-} )$ 
  interacting with the following
  elements: a magnetic field $B$  with  quadripotentiel 
  $A^{mu}=(0,0,B_x,0)$ and a scalar $\sigma$, isoscalar-vector 
  $\omega_\mu$ 
  and isovector-vector $\rho_\mu$, meson fields and leptons
  $(l=e^-, \mu^-)$,   the  Lagrangian density  is expressed as 
  \cite{Typel}:
 \begin{eqnarray}
   \mathcal{L}=  \sum_{b} \bar{\psi_b}\Bigg[  \gamma_\mu(i\partial^\mu-q_bA^\mu-
   g_{\omega b}\omega^\mu-\frac{1}{2}g_{\rho b}\tau.\rho^\mu) \Bigg] \psi_b
   +\sum_{l} \bar{\psi_l}\Bigg[  \gamma_\mu(i\partial^\mu-q_lA^\mu)-
  m_l \Bigg] \psi_l+\\\nonumber
  +\frac{1}{2}\Big( (\partial_\mu\sigma)^2 -m_\sigma^2\sigma^2 \Big) -V(\sigma)-\frac{1}{4}F_{\sigma\lambda}F^{\sigma\lambda}+
  \frac{1}{2}m_\omega^2\omega^2-\frac{1}{4}\omega_{\sigma\lambda}\omega^{\sigma\lambda}-\frac{1}{4}\rho_{\sigma\lambda}\rho^{\sigma\lambda}+
  \frac{1}{2}m_\rho^2\rho_\mu^2,
  \end{eqnarray}
 
where the strong interaction couplings 
$g_{b\sigma}$, $g_{b\omega}$  and  $g_{b\rho}$ 
depend on the density. Furthermore, the following
relations 
$\rho_{\mu\nu}=\partial_\mu\rho_\nu-\partial_\nu\rho_\mu$, 
$F_{\mu\nu}=\partial_\mu A_\nu-\partial_\nu A_\mu$ define 
 the mesonic and electromagnetic field strength,  respectively. 
 We also assume the frozen-field configurations of electromagnetic field
  and neglect the anomalous  magnetic moments of baryons 
  and lepton because their effect is very small. 
  The terms of strong interaction are 
  parametrized by  \cite{Artyom}
  
 \begin{eqnarray}
  g_j(\rho)=a_jg_{j0}\frac{1+b_j(x+d_j)^2}{1+c_j(x+d_j)^2},
 \end{eqnarray}
where $x=\rho/\rho_0$ and  $a_j$, $b_j$, $c_j$,$d_j$ are constants 
(see \cite{Typel} ). The correspondent isovecteur field is also given by
\begin{eqnarray}
  g_{b\rho}=g_{b0}\exp[-a_\rho(x-1)].
 \end{eqnarray}
The mean field approximation constrains the mesonic fields in the 
following equations \cite{Artyom}
\begin{eqnarray}\label{champ moy}
 m_\sigma^2\sigma+\frac{dV}{d\sigma}= \sum_{b} g_{b\rho}n^s_b, \quad m^2_\omega\omega_0=\sum_{b} g_{b\rho}n_b, \quad
 m^2_\omega\rho_{03}=\sum_{b} g_{\rho b}n_b. 
\end{eqnarray} 
With $\sigma$, $\omega^0$, $\rho^0$,
the expectation values of meson fields in the uniform matter, 
$n^s_b$ and $n_b$ the associated scalar and vector 
baryon number densities, respectively.
One recalls the very used scalar field potential  
\begin{eqnarray}
V(\sigma)=\frac{1}{3}p m_N( g_{\sigma N}\sigma)^3+\frac{1}{4}q( g_{\sigma N}\sigma)^4, 
\end{eqnarray}
 where $p$ and  $q$ are dimensionless constants and the values 
 of nucleon-meson couplings and parameters $p$ and $q$ for different 
 models are given  in \cite{Artyom}. \par
  The energy spectra of charged and neutral leptons and  baryons 
  with effective mass $m^*_b=m_b-g_{\sigma b}\sigma$ results from 
  Dirac equation  and can be  expressed by  
\begin{eqnarray}
 E^b_\varUpsilon=(k_z^2+m_b^{*2}+2\varUpsilon |q_b|B)^{\frac{1}{2}}+g_{\omega b}\omega^0+\tau_{3b}g_{\rho b}\rho^0+\Sigma_0^R,\\
 E^b=(k^2+m_b^2)^{\frac{1}{2}}+g_{\omega b}\omega^0+ \Sigma_0^R,\\
 E^l_\varUpsilon=(k_z^2+m_l^{2}+2\varUpsilon |q_l|B)^{\frac{1}{2}}.
\end{eqnarray}
Here, $\tau_{3b}$ is isospin projection, 
the number   $\varUpsilon=n+1/2-sgn(q)/2$ 
represents the Landau levels of the  fermions with 
charge $q$ and spin number $s=\pm 1$. 
Furthermore,  $g_\varUpsilon= 1;2$ correspond to the spin degeneracy
for lowest and others levels of 
Landau, respectively while the last term $\Sigma_0^R$ stays 
for the rearrangement self-energy term and is defined by 

\begin{eqnarray}
 \Sigma_0^R=-\frac{\partial\ln g_{\sigma N}}{\partial  n}m^2_\sigma\sigma^2+
 \frac{\partial\ln g_{\omega N}}{\partial  n}m^2_\omega\omega_0^2+ \frac{\partial\ln g_{\rho N}}{\partial  n}m^2_\rho\rho_0^2,
\end{eqnarray}
with  $n=\Sigma_b n_b$.
We define the scalar densities for neutral 
and charged baryons \cite{Artyom2}  by 
\begin{eqnarray}
 n^{s,n}_b&=&\frac{m_b^{*2}}{2\pi^2}\Big( E^b_fk^b_f- m_b^{*2}\ln|\frac{k^b_f+E^b_f}{m_b^{*}}| \Big),
 \end{eqnarray}
 for neutral baryons with 
 Fermi energy $E^b_f=(k_f^2+m_b^{*2})^{\frac{1}{2}}$  and
\begin{eqnarray}
 n^{s,c}_b&=&\frac{m_b^{*}|q_b|B}{2\pi^2}\sum_\varUpsilon g_\varUpsilon \ln|\frac{k^b_{f,\varUpsilon}
 +E^b_f}{\sqrt{m_b^{*2}+2\varUpsilon|q_b|B}}|,
 \end{eqnarray}
 for charged baryons with Fermi energy 
 $(E^b_f=k_f^2+m_b^{*2}+2\varUpsilon |q_b|B)^{\frac{1}{2}}$. 
The vector densities for both types of baryons are defined by 
 \begin{eqnarray}
  n_b&=&\frac{1}{3\pi^2}k^{b3}_f:\text{neutral baryons },\\
  n_b&=& \frac{|q_{b,l}|B}{2\pi^2}\sum_\varUpsilon g_\varUpsilon k^{b,l}_{f,\varUpsilon}:\text{for charged baryons}.
 \end{eqnarray}
We recall here that  the summation on $\varUpsilon$ 
terminates at the values $\varUpsilon_{\text{max}}$ 
where the square is still positive.
For the strong magnetic field $B\sim 10^{18}G$, 
only few Landau levels are occupied.
For hyperon-meson couplings there are no well-defined rule. 
One can use for these constants quark counting rule \cite{Dover,Schafner}:

 \begin{eqnarray}
  g_{\omega\Lambda}=g_{\omega\Sigma}=\frac{2}{3}g_{\omega N}, \qquad  g_{\omega\Xi}=\frac{1}{2}g_{\omega N},
  \end{eqnarray}
and also
 \begin{eqnarray}
  g_{\rho\Sigma}=2g_{ \rho N}, \qquad  g_{ \rho \Xi}=g_{ \rho N}.
  \end{eqnarray}
The hyperon-meson coupling are assumed to be fixed 
fraction of nucleon-meson couplings explicitly
$g_{iH}=x_{iH}g_{iN}$
where $x_{\sigma H}=x_{\rho H}=0.600$ and  $x_{\omega H}=0.653$ (see \cite{Rabhi}).
The chemical potentials of baryons and leptons are defined by
\begin{eqnarray}
 \mu_b=E^f_b+g_{\omega b} \omega_0+g_{\rho b}\tau_{3b}\rho_0+ \Sigma_0^R, \qquad \mu_l=E^f_l
\end{eqnarray}
 The following conditions should be imposed on the matter 
 for obtaining the EOS:\\
 (i) conservation of baryon number:
 \begin{eqnarray}
  n=\sum_b n_b,
 \end{eqnarray}
(ii) charge neutrality: 
\begin{eqnarray}
  \sum_jq_j n_j=0, \quad j=b,l,
 \end{eqnarray}
 (iii)  beta-equilibrium conditions: 
 \begin{eqnarray}\label{condi}
  \mu_n=\mu_\Lambda=\mu_{\Xi^0}=\mu_{\Sigma^0}, \quad \mu_p=\mu_{\Sigma^+}=\mu_n-\mu_e,\quad 
  \mu_{\Sigma^-}=\mu_{\Xi^-}=\mu_n+\mu_e, \quad \mu_\mu=\mu_e.
 \end{eqnarray}
At given $n$,  equations (\ref{champ moy}) and  (\ref{condi}) 
can be numerically solved and the Fermi energy for the particles 
and the meson 
fields can be found. 
One can also put the correspondingly energy density of 
the matter in the following form 
\begin{eqnarray}
 \varepsilon_m=\sum_b\varepsilon_b+ \sum_l\varepsilon_l+\frac{1}{2}m^2_\sigma+\frac{1}{2}m^2_\omega
 +\frac{1}{2}m_\rho \rho^2_0+W(\sigma).
\end{eqnarray}
The energy densities of neutral and charged baryons are 
respectively given by 
\begin{eqnarray}
 \varepsilon_b^n &=&\frac{1}{4\pi^2}\Big[  k^b_f(E^b_f)^3-\frac{1}{2}m^*_b \Big(  m^*_b k^b_fE^b_f+
 m^{*3}_b \ln|\frac{k^b_f+E^b_f}{m^*_b}|  \Big)     \Big],\\
 \varepsilon_b^c &=& \frac{|q_b|B}{4\pi^2}\sum_\varUpsilon g_\varUpsilon
 \Big(k^b_{f,\varUpsilon}E^b_f +(m_b^{*2}+2\varUpsilon|q_b|B)\ln|\frac{k^b_{f,\varUpsilon}
 +E^b_f}{\sqrt{m_b^{*2}+2\varUpsilon|q_b|B}}\Big)
\end{eqnarray}
The last energy density can also stay for 
leptons by setting $m^*_b\rightarrow m_l$. 
The matter pressure can  be defined by 
\begin{eqnarray}
 p_m= \sum_bp_b+ \sum_lp_l-\frac{1}{2}m^2_\sigma+\frac{1}{2}m^2_\omega
 +\frac{1}{2}m_\rho \rho^2_0-W(\sigma)+ \Sigma_0^R,
\end{eqnarray}
and as consequence, it can be established 
for neutral and charged baryons respectively as  
\begin{eqnarray}
 p_b^n &=&\frac{1}{12\pi^2}\Big[  k^b_f(E^b_f)^3-\frac{3}{2}m^*_b \Big(  m^*_b k^b_fE^b_f+
 m^{*3}_b \ln|\frac{k^b_f+E^b_f}{m^*_b}|  \Big)     \Big],\\
 p_b^c &=& \frac{|q_b|B}{12\pi^2}\sum_\varUpsilon g_\varUpsilon
 \Big(k^b_{f,\varUpsilon}E^b_f +(m_b^{*2}+2\varUpsilon|q_b|B)\ln|\frac{k^b_{f,\varUpsilon}
 +E^b_f}{\sqrt{m_b^{*2}+2\varUpsilon|q_b|B}}\Big)
\end{eqnarray}
Indeed,  the equation of 
state taking into account the magnetic field is given by
\begin{eqnarray}
 \varepsilon=\varepsilon_m+\frac{B^2}{8\pi}, \quad p=p_m+ \frac{B^2}{8\pi}.
\end{eqnarray}
For the models where the magnetic field only depends  on 
the baryon density,
 the parametrization given by \cite{Rabhi2, Ryu} is adopted
\begin{eqnarray}
 B=B_s+B_0[1-\exp(-\beta(n/n_s)^\gamma)],
\end{eqnarray}
$B_s$  is the magnetic field at the surface taken equal to $10^{15} G$ 
in accordance with
the values inferred from observations and 
$B_0$ represents the magnetic field for large densities.
The parameters $\beta$ and
$\gamma$ are chosen in such a way that 
the field decreases fast or slow with the density from the centre to
the surface. In  several
works, it is  used two sets of values: a slowly varying field with
$\beta=0.05$  and $\gamma=2$, and a fast varying  defined by
$\beta=0.05$ and $\gamma=2$. Frequently the magnetic field is expressed 
in units of the critical field $B_e = 4.414 \times 10^{13} G$, 
so that $B = B^{*} B_e$. But in Ref. \cite{Broderick}, it is clearly 
shown that the  structure of a magnetized neutron
stars will be mostly affected by contributions from the magnetic 
field stress 
$P_f = B^2 /8\pi = 4.814 \times 10^{-8}B^{*2}$ Mev f$\text{m}^{-3}$ 
which greatly exceeds the matter pressure $P_m$ at all 
relevant densities for $B^{*}\geq 10^{5}$.  We will use this result in 
our present work in order to investigate the effects of strong magnetic 
field
on the mass-radius relation of neutron via the corrections  of 
Teleparallel. 
\section{Main results}
The effect of strong magnetic field has been developed in
other kind of modified theories of gravity, namely the so 
called $f(R)$ gravity 
\cite{Artyom, Artyom2, Capozziello}. In our case, 
we  also investigate the effect of the strong magnetic field on 
some observables 
features of neutron stars  such as the mass-radius relation. In order to 
reach our goal, we consider the perturbative 
TOV equations and the equation of 
state described in the previous section. For simplicity we 
make use  of the result from 
Ref. \cite{Broderick} notified at the end of the previous section.  
For different values of 
magnetic field parameter $B^{*}$, we plot the mass-radius relations 
for different corrections to Teleparallel term.

\subsection{Quadratic scalar torsion corrections} 
\begin{figure}[htbp]
  \begin{center}
  \includegraphics[width=8 cm]{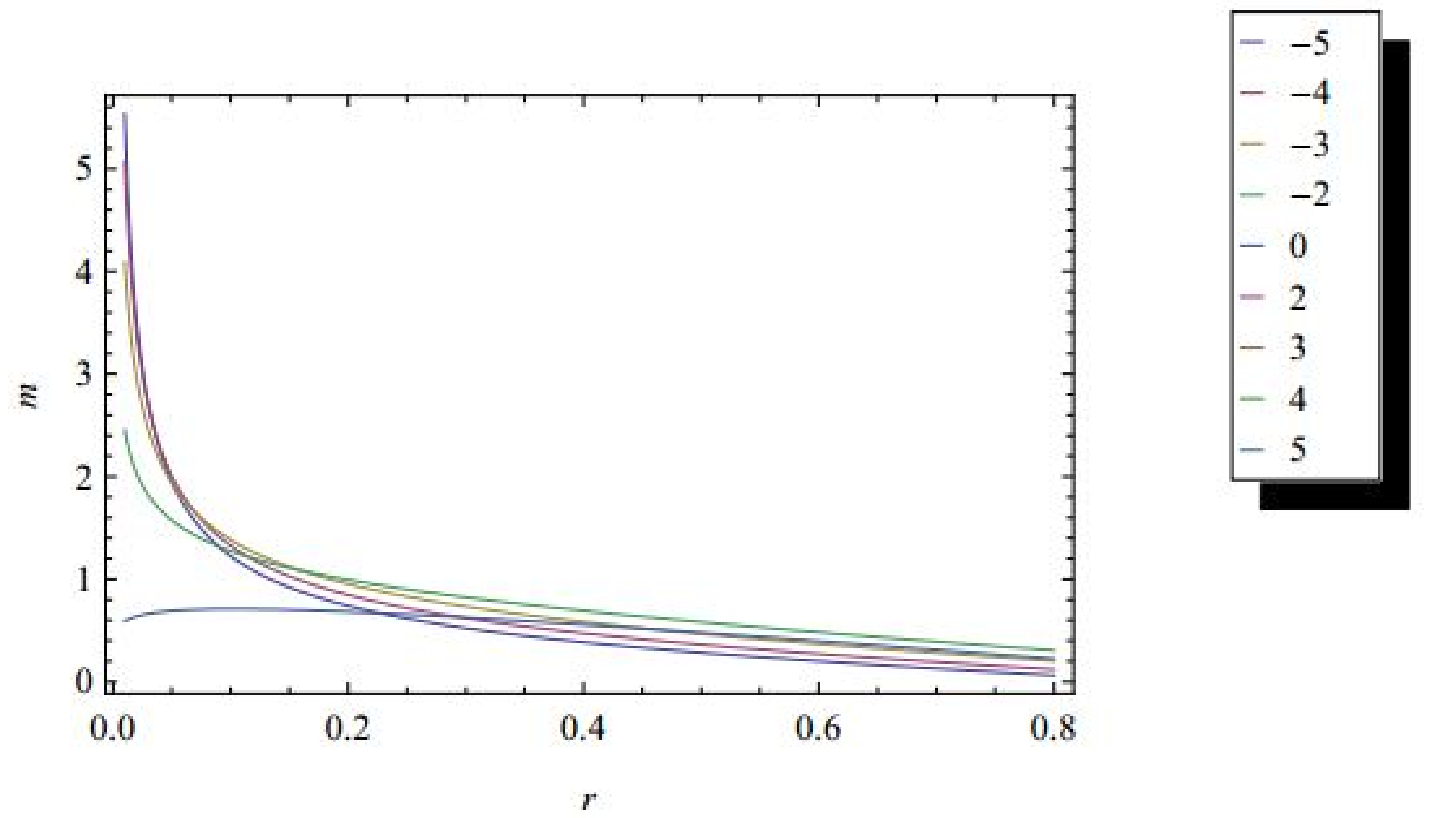}
  \end{center}
  \caption{ The figure shows the mass-radius relation in model $f(T)=T+\xi T^2$  for $B^{*}=0$. 
  The graphs are  plotted for $T^{(0)}=-6\times0.10^{2}$km$^2$.s$^{-2}$.Mpc$^{-2}$ and $\xi=a\times 10^{1}$, with $a=-5,-4,-3,-2,0,2,3,4,5.$}
  \label{fig1}
  \end{figure}

\begin{figure}[htbp]
  \begin{center}
  \includegraphics[width=8 cm]{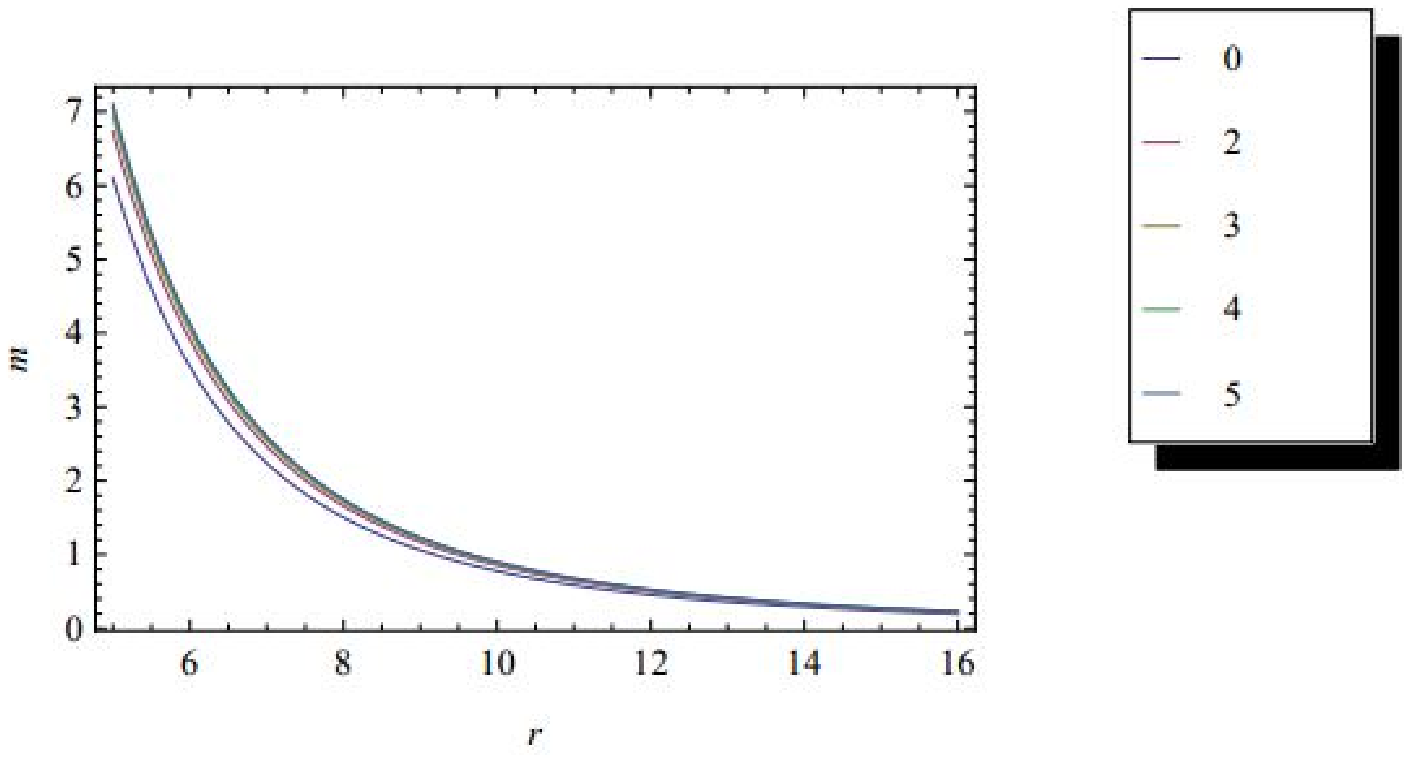}
  \end{center}
  \caption{ The figure shows the mass-radius relation in model $f(T)=T+\xi T^2$  for $B^{*}=5\times 10^{18}$. 
  The graphs are  plotted for $T^{(0)}=-6\times30^{2}$km$^2$.s$^{-2}$.Mpc$^{-2}$  and $\xi=a\times 10^{31}$, with $a=0,2,3,4,5.$}
  \label{fig2}
  \end{figure}

\begin{figure}[htbp]
  \begin{center}
  \includegraphics[width=8 cm]{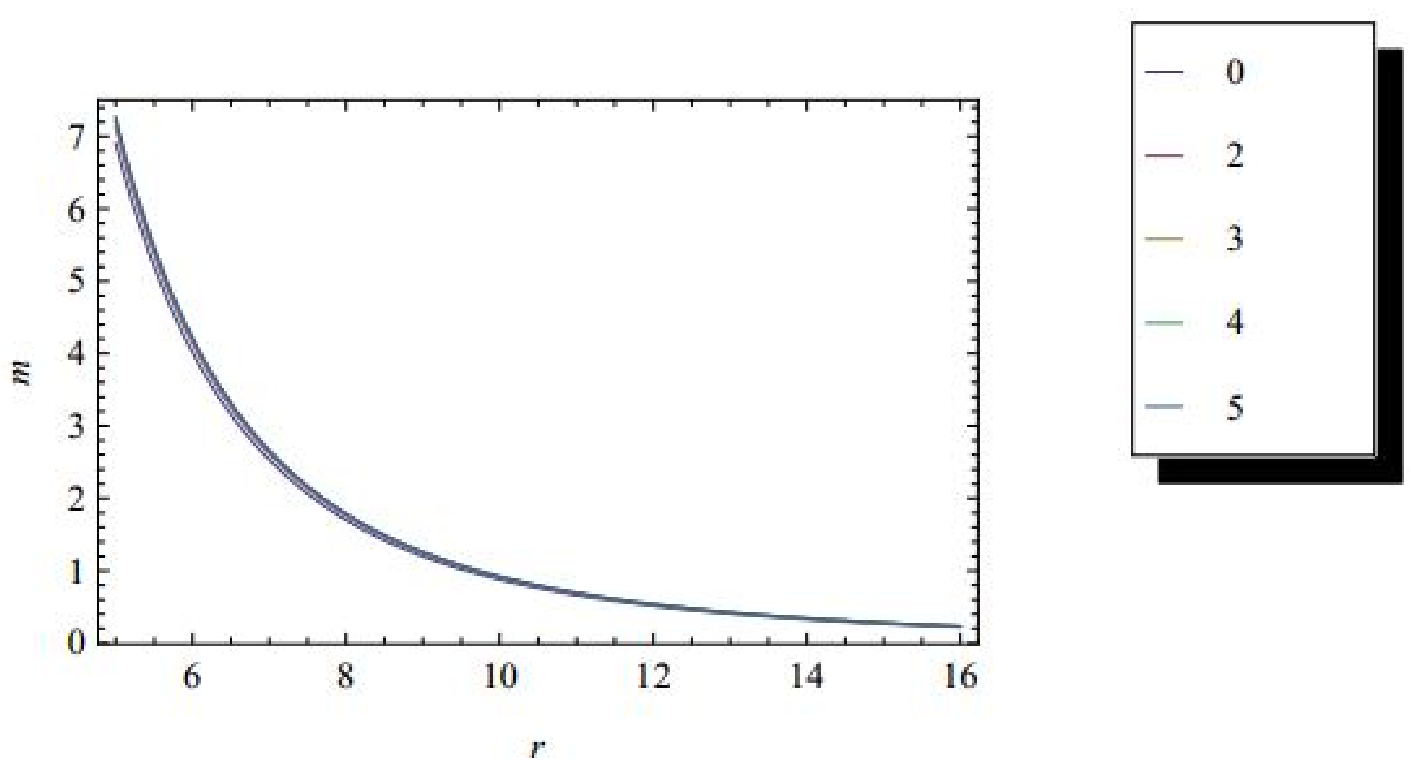}
  \end{center}
  \caption{ The figure shows the mass-radius relation in model $f(T)=T+\xi T^3$  for $B^{*}=3\times 10^{18}$. 
  The graphs are  plotted for $T^{(0)}=-6\times30^{2}$km$^2$.s$^{-2}.$Mpc$^{-2}$  $\xi=a\times 10^{31}$, with $a=0,2,3,4,5.$}
  \label{fig3}
  \end{figure}

Such gravity models are given by 
\begin{equation}
 f(T)=T+\xi T^2.
\end{equation}
It is one case of power-law action function $f(T)=T+b_1T^n$ studied 
by \cite{Houndjo} to explore the mass-radius relation of neutron and
quarks stars 
within $f(T)$ theory. By considering this model, we numerically solve 
the perturbed TOV equations. 
From figure \ref{fig1}, we can note the following features for $B^{*}=0$:
the mass of the neutron stars is very affected by $\xi$ and can 
differ from  the one  in Teleparallel. But for the two figures 
\ref{fig2}  and  \ref{fig3}  which correspond to the large magnetic fields,

the maximal mass of neutron stars increases and can exceed the limit 
value $3M_\odot$ as it is notified in \cite{Artyom}. This can lead to 
stable configuration of neutron stars \cite{Broderick}. 
In this case of large magnetic field, the deviation 
from the Teleparallel is very 
appreciable in the case of  figure \ref{fig2} and 
it requires a large value of the coefficient of corrections.  
 \subsection{Cubic  scalar torsion corrections}
\begin{figure}[htbp]
  \begin{center}
  \includegraphics[width=8 cm]{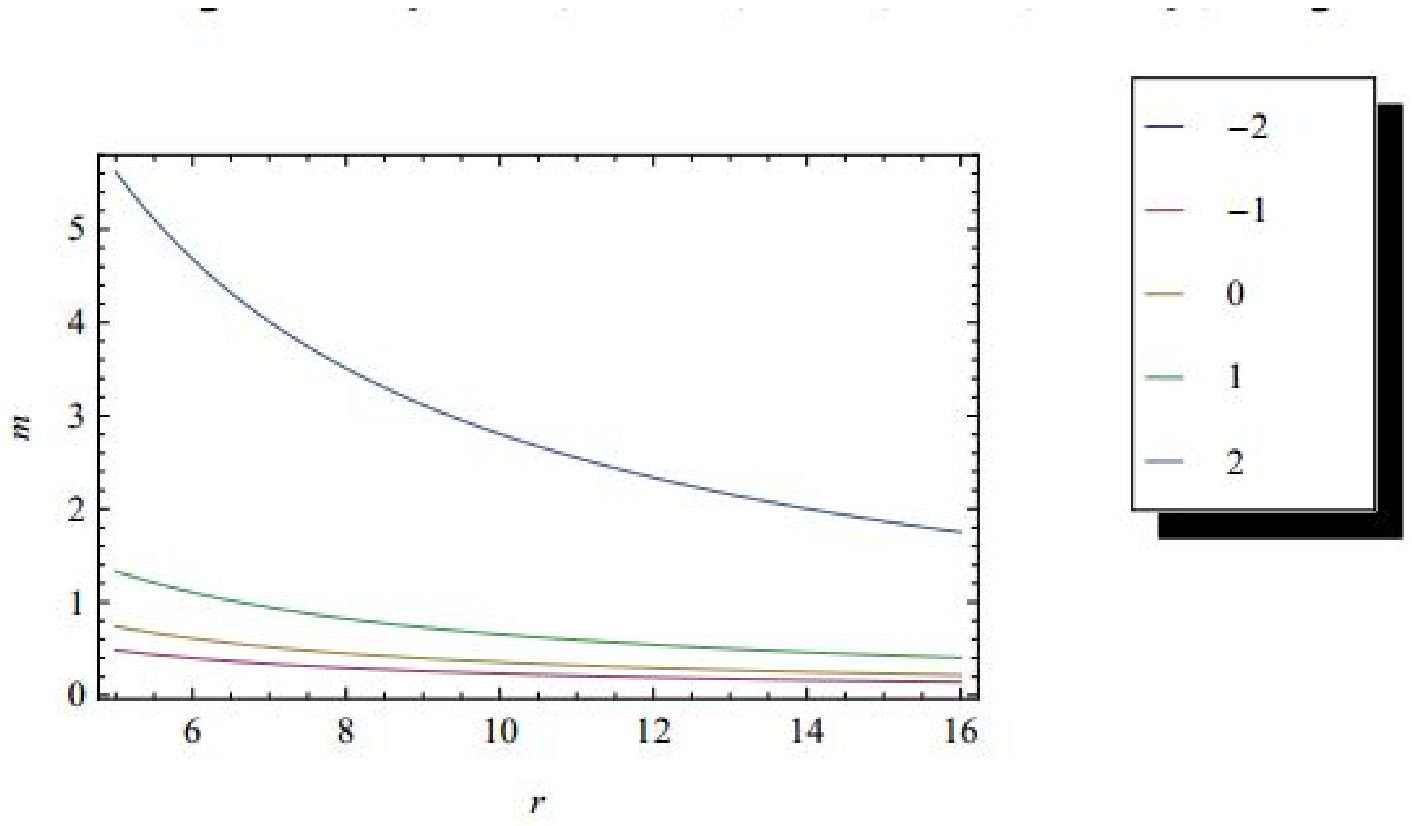}
  \end{center}
  \caption{ The figure shows the mass-radius relation in model $f(T)=T+\xi T^3$  for $B^{*}=0$. 
  The graphs are  plotted for $T^{(0)}=-6\times30^{2}$km$^2$.s$^{-2}$.Mpc$^{-2}$  and $\xi=a\times 10^{3}$, with $a=-2,-1,0,1,2.$}
  \label{fig4}
  \end{figure}

\begin{figure}[htbp]
  \begin{center}
  \includegraphics[width=8 cm]{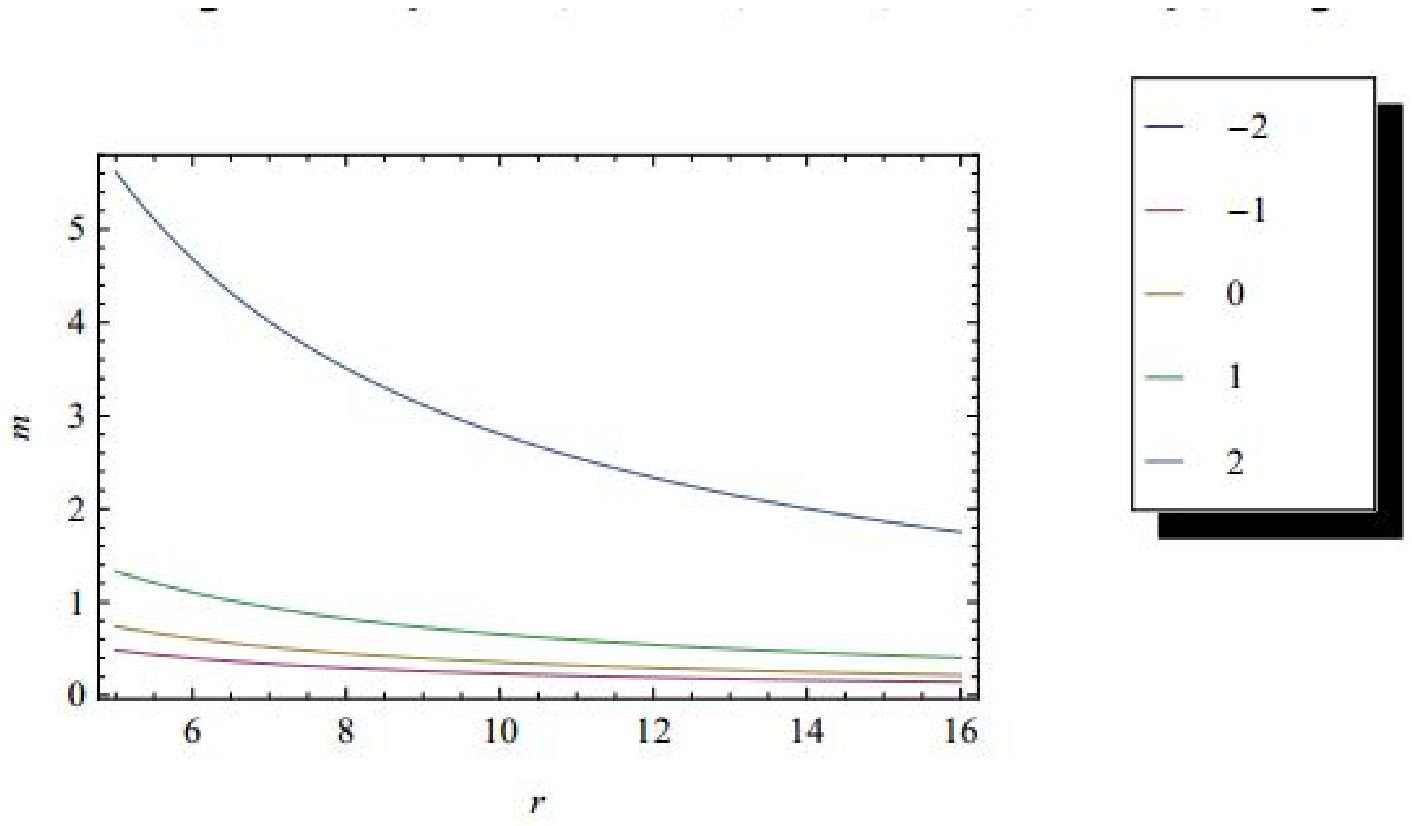}
  \end{center}
  \caption{ The figure shows the mass-radius relation in model $f(T)=T+\xi T^3$  for $B^{*}=5\times 10^{15}$. 
  The graphs are  plotted for $T^{(0)}=-6\times30^{2}$km$^2$.s$^{-2}$.Mpc$^{-2}$  and $\xi=a\times 10^{20}$, with $a=0,2,4,5.$}
  \label{fig5}
  \end{figure}

 The cubic corrections of Teleparallel can be expressed by the model: 
 \begin{equation}
 f(T)=T+\xi T^3,
\end{equation}
 which is also  power-law function. The mass-radius relation according 
 to this model has practically the same behaviours as in the case of 
 quadratic corrections. The interesting feature which appears here is 
 that the deviation from Teleparallel is very clearly appreciable 
 (see  figures \ref{fig4} and \ref{fig5}). In figure \ref{fig4}, 
 the mass of neutron stars increases as the correction 
 parameter $\xi$ evolves
 contrarily to the case  of figure  \ref{fig5}. \par
 In general for the two models, we can note that the mass of neutron
 stars increases due to  the very strong magnetic field which  leads 
 to the  maximum mass configuration of stars as shown in \cite{Rabhi}. 
 These values should be able to describe highly
massive compact stars, such as the one associated to the
millisecond pulsars PSR B1516 $+$ 02B \cite{Paulo}, and the one
in PSR J1748-2021B  \cite{Freire} if the existence of these stellar objects
 is confirmed. 
Due to the magnetic field pressure, the quarks  are  not favored, 
and stars with strong magnetic field are mainly
hadronic with very small hyperon contributions. In addition, according 
to the  considered magnetic field values,
these curves can predict  
stars with very high masses and radius  and
masses of observed compact stars as it was mentioned in \cite{Rabhi2}.
 
 \section{ Conclusion}
  We have studied the influence of a static very  strong  magnetic field  
  on the neutron stars through the power-law $f(T)$ gravity models. 
  Our main goal consists to search  how the strong magnetic field 
  affects the mass-radius relation of neutron star namely the evolution 
  of the 
  neutron stars mass as it is done in  most of GR modify
  theories of gravity such as $f(R)$ theory. In order to reach this goal,
  we have 
  in the first time
   revisited the geometrical equivalence between Teleparallel Theory 
   and GR.  The second step of this work concerns the 
   establishing of TOV equations in the framework of $f(T)$ gravity before 
   proceeding to  its perturbation. We then recall according to the
   literature,
    the equation of the state describing dense matter in magnetic field 
    using a model with baryon octet interacting.
    This leads to the matter equation of state.
    We assume two cases of corrections to Teleparallel in order to solve 
   numerically the perturbed equations and find out  the evolution of 
   the mass-radius relation and the deviation from Teleparallel 
 by considering the very strong magnetic field. 
  Our investigations show that the mass of neutron stars 
  increase even in the case of null magnetic field where the correction 
  parameter $\xi$ plays an important role.  For the both  quadratic and 
   cubic  corrections, the mass of neutron stars can increase  and the 
   deviation from Tele-Parallel is very  appreciable in the case
of cubic corrections. Moreover the evolution of the mass is very 
specified in this case: the mass increase as $\xi$ evolves
( figure \ref{fig4})  and decreases in case of figure \ref{fig5}.  
Due to the consideration of  ultrastrong magnetic , the magnetic 
pressure has only been considered  and by consequence the quark 
star is not favored whereas the hadronic with very 
small hyperon contributions can be compatible with the results 
obtained in this work. 
 \vspace{2cm}

\begin{center}
 \rule{8cm}{1pt}
\end{center}

 \end{document}